\documentclass[%
 aip,
 amsmath,amssymb,
 reprint,%
]{revtex4-1}

\usepackage{graphicx}
\usepackage{dcolumn}
\usepackage{bm}

\usepackage[utf8]{inputenc}
\usepackage[T1]{fontenc}
\usepackage{mathptmx}
\usepackage{etoolbox}

\makeatletter
\def\@email#1#2{%
 \endgroup
 \patchcmd{\titleblock@produce}
  {\frontmatter@RRAPformat}
  {\frontmatter@RRAPformat{\produce@RRAP{*#1\href{mailto:#2}{#2}}}\frontmatter@RRAPformat}
  {}{}
}%
\makeatother
\begin{document}

\preprint{AIP/123-QED}

\title[]{The effect of doping layers position on the heterojunction sharpness in (In,Al)As/AlAs quantum dots}

\author{T. S. Shamirzaev}
\email{tim@isp.nsc.ru}
\affiliation{Rzhanov Institute of Semiconductor Physics\unskip, Siberian Branch of the Russian Academy of Sciences\unskip, 630090\unskip, Novosibirsk\unskip, Russia}
\author{D. R. Yakovlev}
\affiliation{Experimentelle Physik2\unskip, Technische Universit{\"{a}}t Dortmund\unskip, 44227\unskip, Dortmund\unskip, Germany}
\affiliation{Ioffe Institute, Russian Academy of Sciences, 194021, St. Petersburg, Russia}
\author{M. Bayer}
\affiliation{Experimentelle Physik2\unskip, Technische Universit{\"{a}}t Dortmund\unskip, 44227\unskip, Dortmund\unskip, Germany}
\affiliation{Research Center FEMS, Technische Universit\"at Dortmund, 44227 Dortmund, Germany}

\date{\today}

\begin{abstract}
Effect of doped layer placed in structures with indirect band-gap (In,Al)As/AlAs quantum dots (QDs) on heterointerface sharpness is investigated. We demonstrate that growth of $n$ ($p$) doped layer below QDs sheet leads to pronounced deceleration (acceleration) for dynamics of exciton recombination (which is very sensitive to heterointeface structure in these QDs) in compare with the undoped structure. Opposite, the placing of the same doped layers above the QDs sheet does not effect on the exciton recombination dynamic at all. The experimental data are explained by increase (decrease) charged vacancy formation rate in the cation sublattice, that result in QD/matrix interface bluring (sharping), with the increases in the electron (hole) concentration at this heterointerface formation. The thicknesses of the diffuse layer on  QD/matrix heterointerface estimated is in range from 0 up to 5 in the lattice constant depending on doped layer placing.
\end{abstract}

\maketitle

Low-dimensional semiconductor heterostructures are  interesting  from  the  viewpoint  of  both  basic  physics and  potential applications~\cite{Orton,Mounika,Klingshirn,Rafailov}. One of the advantages of heterostructures fabricated by molecular-beam epitaxy technique is a heterointerface sharpness that is very important to provide high electron mobility in microwave devices or to develop ultra-low-loss crystalline microresonators for quantum cascade lasers in THz range~\cite{Zhu,Poornachandran,Xu,Consolino}. To grown the structurally perfect  heterointerface  a lot of  technologies have been developed based on tracking epitaxy parameters that determine the diffusion and incorporation of atoms on the growth surface (such as, a substrate temperature, the rate of atomic deposition, and the ratio of atomic flows during the growth of composite compounds by the molecular-beam epitaxy)~\cite{Alam,Book1,Mu,Jasik}. However, these degrees of freedom are often not enough to overcome all challenges that arise at the fabrication of semiconductor heterostructures. Additional possibilities for controlling growth processes appear when one  takes into account the  electronic subsystem, which can change under the influence of external factors such, for instance, as the  electron-hole pairs generation under illumination of the structure during the growth process~\cite{Alberi,Park,Park1,Zhu5,Sanders}, and also as a result of doping various layers of the structure~\cite{Schubert}. Indeed, atoms may diffuse within epitaxial layers and across heterojunction aided by vacancies. Since vacancies can be in different charge states~\cite{Seebauer},  low or high values of the Fermi level provided by illumination or doping of layers in  epitaxy process can reduce or increases the vacancies formation energy~\cite{Lee,Shamirzaev13n,Shamirzaev13n1}, as a result, heterojunction sharpness can be changed.  However, the effect of a doping layer position inside growing heterostructure on heterointerface sharpness  has been scarcely studied so far.
We demonstrated recently that indirect band-gap (In,Al)As/AlAs QDs with type-I band alignment are a very good model object to the heterointerface sharpness verification. Radiative recombination time of indirect band gap exciton in these QDs are very sensitive to the local thickness of the diffused (In,Al)As layer, which is appeared at the  QD-matrix interface as a result of annealing. It is varying from ten nanosecond up to hundreds microsecond with change heterointerface sharpness~\cite{Shamirzaev84}.

In this paper, we investigate effect of doped layer placed in structure with indirect band-gap (In,Al)As/AlAs quantum dots (QDs) on heterointerface sharpness.  We  demonstrate  that a growth of a $n$ ($p$) heavily doped layer before QDs formation results in decreasing (increasing) of QD/matrix heterointerface sharpness compare to undoped case. Meantime, in the opposite case, when heavily doped  layers  are growing  after the  QDs formation the heterointerface sharpness does not differ from that in the undoped case. Experimental results are explained by acceleration (deceleration) charged vacancy formation rate in the cation sublattice with the increases in the electron (hole) concentration during QD/matrix heterointerface formation. We suggest that this effect is a result of dependence of Schottky charged vacancy formation rate on carrier concentration.

The studied self-assembled (In,Al)As QDs, embedded in an AlAs matrix, were grown by molecular-beam epitaxy on semi-insulating (001)-oriented GaAs substrates.
\begin{figure}[]
\includegraphics* [width=8 cm]{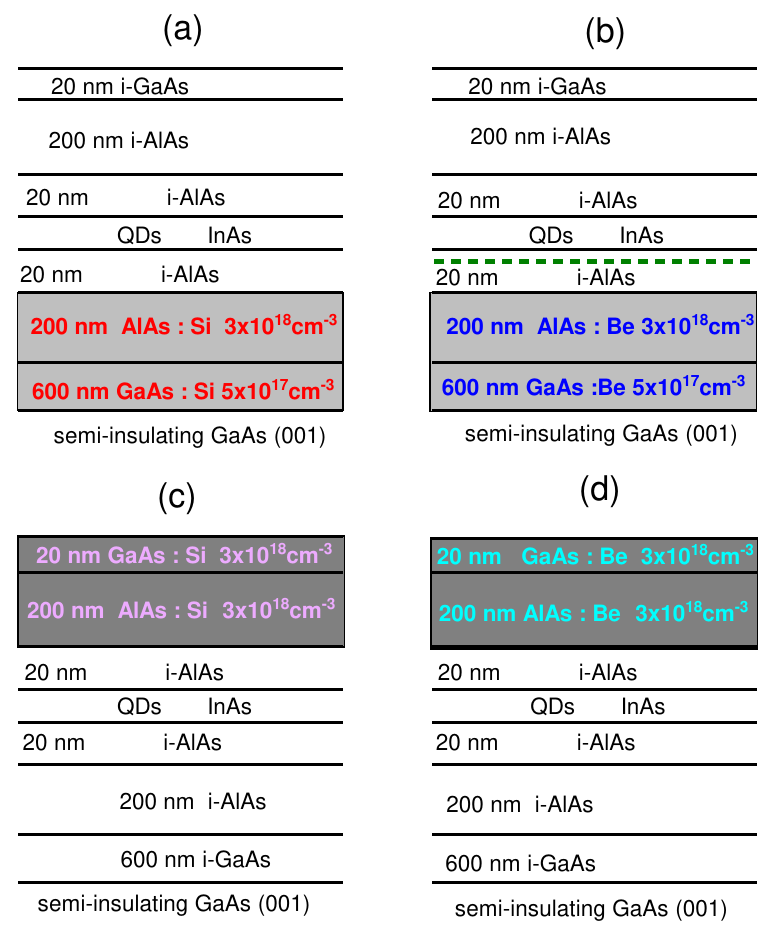}
\caption{\label{fig1} The design of studied heterostructures. Layers were doped by silicon ($n$-type) or beryllium (p-type). Structures with doped layer grown below QDs sheet (a) A and (b) B. Structures with doped layer grown above QDs sheet (c) C and (d) D. Colors of doped layers are selected the same as in Fig.2(b) for PL dynamics measured for correspondent structure. Dashed green line indicates place of additional $p$-doped layer with thickness of 2~nm that placed at 2~nm below QDs sheet in structure F. Color of doped layer symbols in structures  corresponds to the dynamics curve color in Fig.~\ref{fig2} for convenience sake.}
\end{figure}
The designs of studied structures are shown in Fig.~\ref{fig1}. A QDs sheet is sandwiched between AlAs layers grown on top of a 600-nm-thick GaAs buffer layer. Doped layers were grown below (structures A ($n$-doped Si $3 \times 10^{18}$ cm$^{-3}$) and B ($p$-doped Be $3 \times 10^{18}$ cm$^{-3}$ )) and above  (structures C ($n$-doped) and D ($p$-doped)) of a QDs sheet that was sandwiched between undoped AlAs layers with 20-nm thickness. A similar fully undoped structure (E) has been grown  as a reference.  The structure F had the design similar to structure D but with an additional thin $p$-doped ($3 \times 10^{18}$ cm$^{-3}$) layer placed at 2~nm below QDs sheet.  The nominal amount of deposited InAs was about 2.5 monolayers. The QDs were formed at the temperature of 510$^{\circ}$C with a growth interruption time of 30~s.  A 20-nm-thick GaAs cap layer protected the AlAs layer against oxidation. Further details for the epitaxial growth QDs in the AlAs matrix are given in Ref.~\onlinecite{Shamirzaev78}. From the growth conditions and model calculations an average QD composition was estimated as In$_{0.75}$Al$_{0.25}$As~\cite{Shamirzaev78}. The size and density of the lens-shaped QDs were measured by transmission electron microscopy, yielding an average diameter of $13\div15$~nm  and a density of about $2.5 \times 10^{10}$ dots per cm$^{2}$. The relatively low QDs density prevents carrier redistribution between the QDs~\cite{ShamirzaevNT,ShamirzaevSST}.

The samples are immersed in pumped liquid helium. The time-integrated and time-resolved PL measurements were performed at a temperature of $T = 1.8$~K. The PL were exited by the third harmonic of a Q-switched Nd:YVO$_4$ laser (3.49~eV) with a pulse duration of 5~ns. The pulse-repetition frequency was varied from 2 to 100~kHz and the excitation energy density was kept below 100 nJ/cm$^{2}$, which corresponds to about 30$\%$ probability of QD occupation with a single exciton~\cite{Shamirzaev84}. The light emitted in the case measurement of PL dynamics at selected energy was detected by a GaAs photomultiplier operating in the time-correlated photon-counting mode. In order to monitor the PL decay in a wide temporal range up to 350~$\mu$s the time resolution of the detection system was varied between 1.6~ns and 256~ns.

\begin{table*}[]
    \caption{Parameters of the studied (In,Al)/AlAs QDs evaluated from best fit  to the experimental data. $d$ is a thickness of the
    diffused layer formed at the QD-matrix interface in lattice constant unit.}
  \begin{ruledtabular}
        \begin{tabular}{lcccccc}
           Parameter              &         A       &      B           &  C             &       D          &  E             &  F              \\
            Doping                & $n$-type/bottom & $p$-type/bottom  & $n$-type/upper & $p$-type/upper   & undoped        & $p$-type/bottom  \\
            $\gamma$              & $1.6~\pm$0.04   & $1.95~\pm$0.02   & $2.05~\pm$0.01 & $2.07~\pm$0.01   & $2.09~\pm$0.01 & $3.43~\pm$0.01      \\
            $\tau_{0}$            & 1250$~\pm$10~ns & 55$~\pm$4~ns     & 320$~\pm$10~ns & 340$~\pm$10~ns   & 370$~\pm$10~ns & 8$~\pm$1~ns    \\
            $d$                   &      5          &      2           & $3\div4$       & $3\div4$         &  $3\div4$      & 0    \\
        \end{tabular}
  \end{ruledtabular}
    \label{tab:parameters}
\end{table*}

\begin{figure}[t]
\includegraphics* [width=7.0cm]{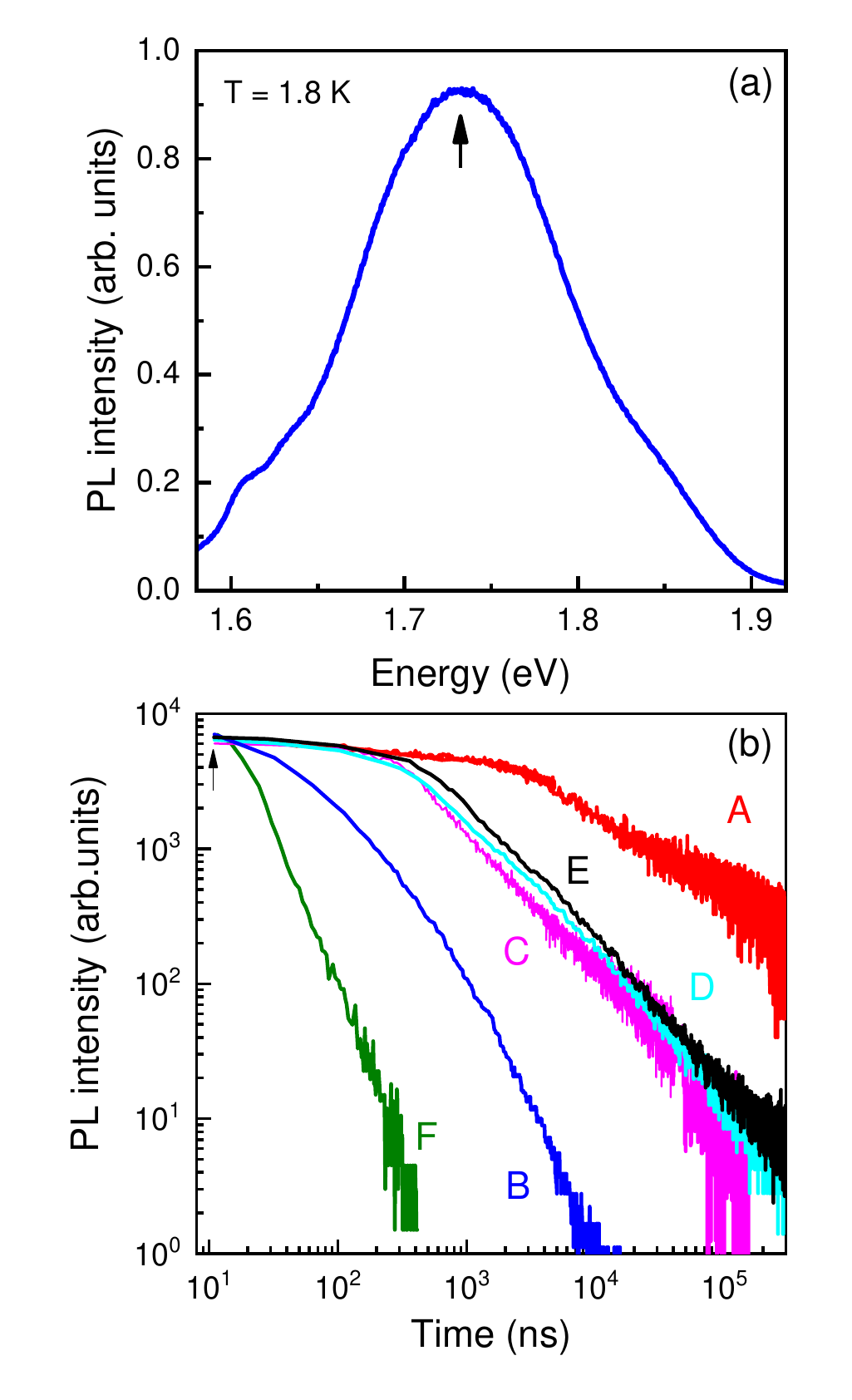}
\caption{\label{fig2} (a) Typical time-integrated PL spectrum for In(Al)As/AlAs QDs. The energy of 1.73~eV, where the PL dynamics is measured, is marked by an arrow;  (b) Normalized dynamics of exciton recombination at maximum of the PL band for undoped structure E (black) as well as for: bottom-layer-doped structures ($p$-doped, B - blue), and ($n$-doped, A - red), top-layer-doped structures ($p$-doped, D - cyan), and ($n$-doped, C - magenta), and structure F with additional $p$-doped layer (green). The arrow marks the end of the excitation pulse.}
\end{figure}
A low temperature time-intergrated PL spectrum ($T=1.8$~K) of  (In,Al)As/AlAs QDs is shown in  Fig.~\ref{fig2}(a). The spectrum shape and energy position do not depend on the doping layers type and placing. The broad PL band of the QD emission has maximum at 1.73~eV and a full width at half maximum of 120~meV, that reflected dispersion of the QD parameters, since the exciton energy depends on the QD size, shape, and composition~\cite{Shamirzaev78,Shamirzaev84}. The energy position of this band evidences that the conduction-band minimum in these QDs belongs to the X-valley, i.e., it is indirect in the momentum space~\cite{Shamirzaev78}.

The dynamics of exciton recombination at maximum of the PL band for different structures are shown in Fig.~\ref{fig2}(b). The transient PL data are plotted on a double-logarithmic scale, which is convenient to illustrate the nonexponential character of the decay over a wide range of times and PL intensities. All dynamics are normalized to the same initial intensity for ease of comparison. For all structures the recombination dynamics demonstrate two stages: (i)  some relatively  flat decay immediately after the excitation pulse,  (ii)    further PL decay that can be described by the power-law function $I(t)\sim (1/t)^\alpha$, as shown in our previous studies~\cite{Shamirzaev78,Shamirzaev84,Ivanov97}. Such dynamics result from the superposition of multiple monoexponential decays with different times varying with the size, shape, and composition of indirect-band-gap QDs.

One can see, that for undoped structure E and structures C and D with doped layers grown above QDs sheet the PL dynamics are very similar. However, the dynamics change drastically for structures with doped layers grown below QDs sheet. For structure with $n$-doped layer, A, PL decay strongly decelerates, while for structures with $p$-doped layer, B and F, the decay accelerate and the more the closer the doped layer is to the QDs sheet.

It was early demonstrated that PL dynamics of QDs ensemble can be described by the following equation~\cite{Nikolaev,Shamirzaev84}:
\begin{equation} \label{eq1}
I(t) = \int_0^{\infty} G(\tau) \exp \left( -\frac{t}{\tau} \right )
d\tau,
\end{equation}
where $G(\tau)$ is distribution function described dispersion of exciton recombination times in indirect band gap QDs ensemble. It can be described as~\cite{Shamirzaev84}:
\begin{equation} \label{eq2}
G(\tau) = \frac{C}{\tau^{\gamma}}
\textnormal{exp}\left[-\frac{\tau_{0}}{\tau}\right].
\end{equation}
Here $C$ is a constant, $\tau_0$ characterizes the maximum of the distribution of exciton lifetimes, and the parameter $\gamma$  is defined as $\alpha$ + 1. The $\gamma$ can be extract directly from the power-law decay (1/$t$)$^\alpha$ represented in Fig.~\ref{fig2}(b).

\begin{figure}[]
\includegraphics* [width=6.0cm]{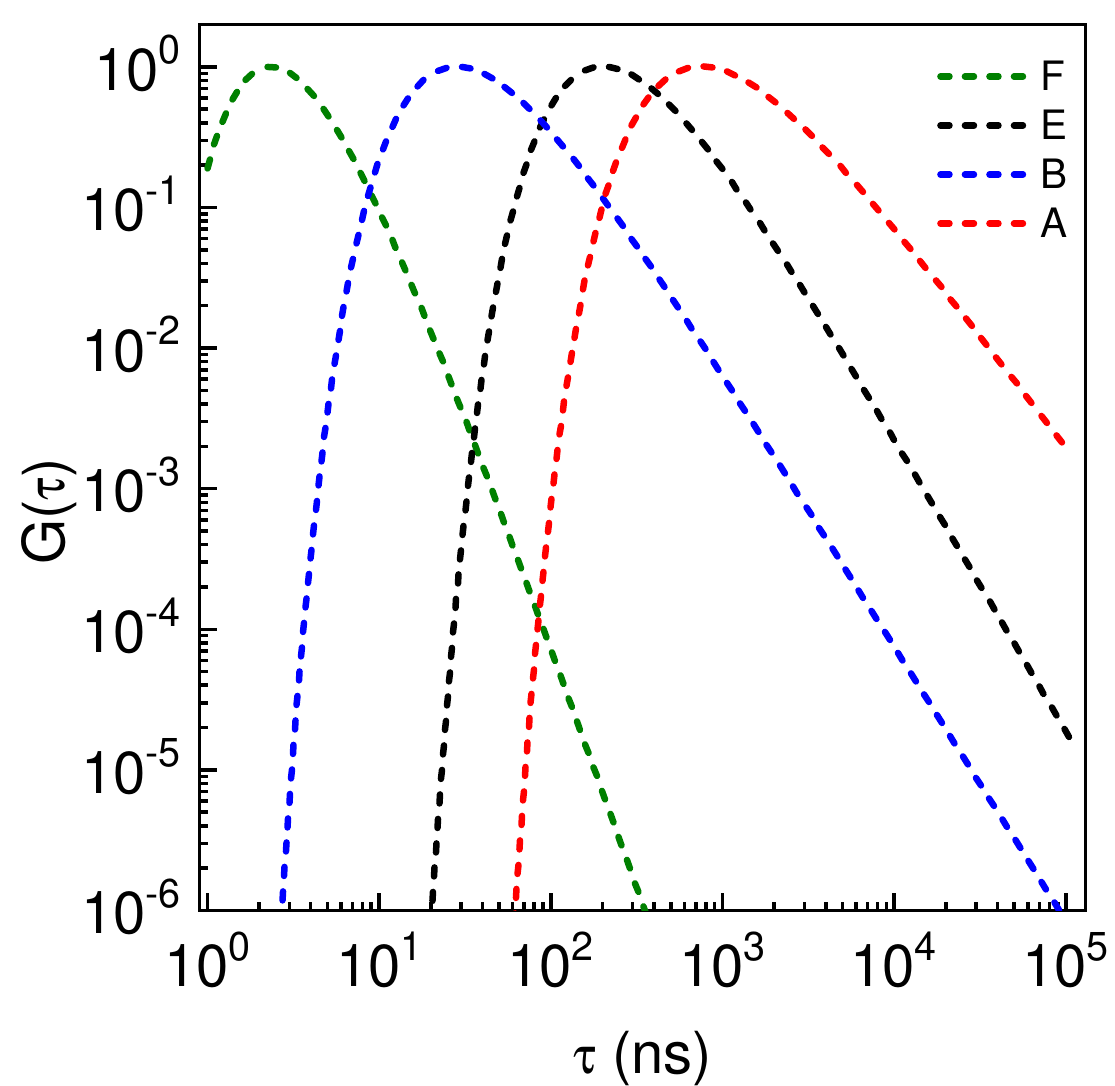}
\caption{\label{fig3} Distributions of exciton lifetimes, $G(\tau)$, corresponding to structures A, B, E, and F obtained by best fit of recombination dynamics.}
\end{figure}

Using the model approach suggested in our recent study~\cite{Shamirzaev84}, we obtained these distribution functions for studied structures by fitting the recombination dynamics. The parameters of best fit are collected in Table~\ref{tab:parameters}  and correspondent distribution function $G(\tau)$ are shown in Fig.~\ref{fig3}. One can see that distribution of exciton lifetime strongly shifts depending on doped layer position. In the cases of undoped structure and for position of doped layer above QDs sheet most probable lifetime $\tau_0 = 320 \div 370$~ns, while for doped layers grown bellow QDs it decreased in the case of $p$-doping $\tau_0=8$~ns (structure F) and  $\tau_0 = 55$~ns (structure B) and it increased $\tau_0 = 1250$~ns (structure A).

As we recently demonstrated theoretically in the Appendix to Ref.~\onlinecite{Shamirzaev84}, the lifetime of indirect in momentum space exciton is determined by momentum scattering at the interface and can be described by the expression $\tau\propto exp(d/a + d/L)$,  where $a$ is the lattice constant, $L$ is the QD height, and $d$ is the thickness of the diffused (In,Al)As layer at the QD-matrix interface. Using this expression we estimate  blurring of heterointerface QD/matrix. Taking into account that an aspect ratio of studied QDs is about $1:3 - 1:4$~\cite{ShamirzaevAPL92} we estimate $L$ as of about 4.0~nm and taken that  $d/L$ is smaller than $a/d$. Therefore, we neglect by $d/L$ contribution and taken $\tau \propto exp(d/a)$. Assuming that in structure F with the fastest exciton lifetime heterointerface is as sharp as possible ($d_{\text{F}}=0$) we roughly estimate thickness of the diffuse layers in the units of lattice constant for all structures from relations between exciton lifetimes as, $d \cong 2a$, $3a\div4a$, and $5a$ for structure B, E, and A, respectively.

We demonstrated recently that intermixing of (In,Al)As/AlAs QDs induced by vacancy-mediated diffusion at post-grown annealing at temperatures significantly exceed  temperature of epitaxy is enhanced (suppressed) in $n$ ($p$) doped heterostructures due to effect of charge carriers sign and concentration on charged Frenkel vacancy formation rate~\cite{Shamirzaev13n}. The result of present study is clearly demonstrated that type and concentration of charge carriers located near growth surface can also effect on formation of heterointerface atomic structure already during an epitaxial growth procedure. We suggest that this effect is a result of dependence of formation rate for charged vacancy  at the surface (Schottky defects) on carrier concentration.

Thus, taking into account high sensitivity of indirect-band-gap exciton radiative lifetime in (In,Al)As/AlAs quantum dots with type I band alignment to heterointerface atomic structure we examined an effect of doped layer position in heterostructures with the QDs on heterointerface sharpness. In has been demonstrated that doped layers grown below QDs sheet effected strongly on sharpness of QD/matrix interface (it is sharping with $p$-doping and blurring with $n$-doping), while doped layers grown above QDs sheet does not effect on the interface sharpness. The blurring degree heterointerface is controlled by vacancy assisted disordering and it depends on charged vacancy formation rate, which is ruled by electron and hole concentrations in the interface region during it formation.

\begin{acknowledgments}
Experimental   activities   conducted   by   T.S.S.  were supported   by a  grant   of   Russian   Science   Foundation   (project   No.   22-12-00022-P), D.R.Y, and M.B. acknowledge the financial support by the Deutsche Forschungsgemeinschaft through the Collaborative Research Center TRR142 (Project A11).
\end{acknowledgments}

\section*{Data Availability}
The data that support the findings of this study are available from the corresponding author upon reasonable request.

\section*{REFERENCES}

\bibliography{InAsAlAsDoped}

\end{document}